\documentstyle[psfig,macros]{mn} 
\begin{document}
%

\title[Generating Merger Trees]
      {Generating Merger Trees for Dark Matter Haloes:\\ A Comparison of Methods}

\author[Jiang \& van den Bosch]
       {Arthur Fangzhou Jiang\thanks{E-mail:fangzhou.jiang@yale.edu} 
        \& Frank C. van den Bosch \\
        Department of Astronomy, Yale University, New Haven, CT 06511, USA}


\date{}

\pagerange{\pageref{firstpage}--\pageref{lastpage}}
\pubyear{2013}

\maketitle

\label{firstpage}


\begin{abstract}
Halo merger trees describe the hierarchical mass assembly of dark matter haloes, and are the backbone for modeling galaxy formation and evolution. Merger trees constructed using Monte Carlo algorithms based on the extended Press-Schechter (EPS) formalism are complementary to those extracted from N-body simulations, and have the advantage that they are not trammeled by limited numerical resolution and uncertainties in identifying (sub)haloes and linking them between snapshots. This paper compares multiple EPS-based merger tree algorithms to simulation results using four diagnostics: progenitor mass function (PMF), mass assembly history (MAH), merger rate per descendant halo, and the unevolved subhalo mass function (USMF).  In general, algorithms based on spherical collapse yield major-merger rates that are too high by a factor of two, resulting in MAHs that are systematically offset. Assuming ellipsoidal collapse solves most of these issues, but the particular algorithm investigated here that incorporates ellipsoidal collapse dramatically overpredicts the minor-merger rate for massive haloes. The only algorithm in our comparison that yields MAHs, merger rates, and USMFs in good agreement with simulations, is that by Parkinson et al. (2008).  However, this is not a true EPS-based algorithm as it draws its progenitor masses from a PMF calibrated against simulations, rather than `predicted' by EPS.  Finally we emphasize that the benchmarks used to test the EPS algorithms are obtained from simulations and are hampered by significant uncertainties themselves. In particular, MAHs and halo merger rates obtained from simulations by different authors reveal discrepancies that easily exceed 50 percent, even when based on the same simulation. Given this status quo, merger trees constructed using the Parkinson et al. algorithm are as accurate as those extracted from N-body simulations.
\end{abstract}




\section{Introduction} 
\label{sec:intro}

Halo merger trees describe the hierarchical mass assembly of dark
matter haloes. They are the backbone for modeling the formation and
evolution of galaxies (see Mo, van den Bosch \& White 2010), and they
are the core ingredient in semi-analytical models that aim to describe
the substructure of dark matter haloes (e.g., Oguri \& Lee 2004;
Zentner \etal 2005; Taylor \& Babul 2005a,b; van den Bosch \etal 2005;
Gan \etal 2010). Two different methods are used to construct halo
merger trees: Monte Carlo methods based on the extended
Press-Schechter (EPS; Bond \etal 1991) formalism and numerical
$N$-body simulations. Although the rapid advances in computer
technology has shifted focus from EPS-based merger trees to extracting
merger trees from numerical simulations (e.g., Kauffmann \etal 1999a,b;
Benson \etal 2000; Helly \etal 2003; Kang \etal 2005; Springel \etal
2005; Han \etal 2012; Behroozi \etal 2013), EPS-based methods
remain an important and powerful alternative for a number of
reasons. 

First of all, EPS methods are typically much faster and
more flexible. Although a cosmological simulation typically yields
merger trees (after analysis) of thousands of haloes at once, whereas
an EPS-based method constructs halo merger trees for each halo at a
time, the limited force resolution and mass resolution of N-body simulations
introduce serious systematics. In particular, the merger trees of
more massive haloes are better resolved, i.e., probe down to
progenitor masses that are a smaller fraction of the mass of the final
host halo. This complicates a proper analysis of how the (statistical)
properties of merger trees scale with halo mass.  

Furthermore,
to explore the dependence on cosmological parameters typically requires
one to run large sets of simulations.  The EPS-based method, on the
other hand, can typically construct halo merger trees at high mass
resolution (i.e., down to progenitors with a mass as small as
$10^{-5}$ times that of the final host halo mass) in a matter of
seconds, and can therefore construct merger trees for large sets of
haloes of different masses and/or cosmologies in a fraction of the
time required to run and analyze a full blown cosmological simulation.

In addition, it is important to realize that although simulations more
reliably capture the physics of gravitational collapse and halo growth
in a hierarchical universe than EPS theory, extracting reliable merger
trees from simulations is subject to a large number of tricky,
systematic issues. In particular, depending on the algorithms used to
identify haloes and subhaloes, and to link haloes between different
snapshots, one can obtain merger trees that differ substantially
(e.g., Harker \etal 2006; Fakhouri \& Ma 2008, 2009; Fakhouri \etal 2010; Genel \etal
2008, 2009, 2010). Indeed, a recent comparison of
ten different merger tree construction algorithms applied to the same
simulation output has revealed a discomforting amount of disparity
(Srisawat \etal 2013).

In this paper, we compare a number of different methods, available in
the literature, that are used to construct EPS-based merger trees.
These cover a variety of strategies: some algorithms make the implicit
assumption that each branching point in the tree represents a binary
merger, while others allow for multiple mergers per branching point;
some algorithms assure that the sum of the progenitor mass exactly
equals the mass of the descendant, while others admit mild violation
of mass conservation; some algorithms are based on the (standard)
spherical collapse model, while others adopt the more realistic
picture of ellipsoidal collapse; and finally, some algorithms are
self-consistent, while others use a progenitor mass function that is
inconsistent with EPS. By comparing with numerical simulations, we
test how accurately these methods can reproduce various statistics of
the hierarchical assembly of dark matter haloes, such as the unevolved
subhalo mass function (i.e., the mass function of subhaloes at
accretion), merger rates, and mass assembly histories.

This paper is organized as follows: \S\ref{sec:mergertrees} discusses
the anatomy of merger trees, and the challenges associated with their
construction using either numerical simulations or semi-analytical
methods based on the excursion set formalism.  \S\ref{sec:algorithms}
describes the various merger tree algorithms that are tested and
compared in terms of their progenitor mass functions
(\S\ref{sec:PMFs}), mass assembly histories (\S\ref{sec:MAHs}),
merger rates per descendant halo (\S\ref{sec:MergerRate}) and their
unevolved subhalo mass functions (\S\ref{sec:USMF}). Results are
summarized in \S\ref{sec:Discussion}.

\section{Halo Merger Trees}
\label{sec:mergertrees}

\subsection{Anatomy of a Merger Tree}
\label{sec:anatomy}

\begin{figure}
\centerline{\psfig{figure=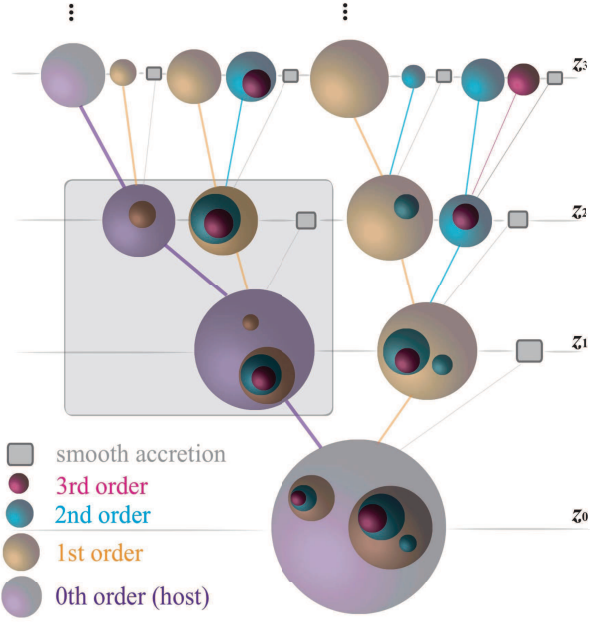,width=\hssize}}
\caption{Illustration depicting the anatomy of a merger tree for a
  host halo (purple sphere at the bottom) at redshift $z=z_0$. The
  purple spheres to the left illustrate the assembly history of the
  main progenitor. We refer to these as `zeroth-order' progenitors,
  and they accrete `first-order' progenitors, which end up as
  (first-order) subhaloes at $z=z_0$. In turn, these first-order
  progenitors accrete second-order progenitors which end-up as
  second-order subhaloes (sub-subhaloes) at $z=z_0$, etc.  The size of a
  sphere is proportional to its mass, while its color reflects its
  order, as indicated.  The large, shaded box highlights a single
  branching point in the tree structure, which shows a descendant halo
  plus its single-time-step progenitors. This is the building block of an EPS merger tree. 
  The small shaded boxes
  present at each branching point reflect `smooth accretion', as
  defined in the text.}
\label{fig:mergertree}
\end{figure}

Before describing how to construct halo merger trees using the EPS
formalism, we first define some terminology used throughout this
paper. Fig.~\ref{fig:mergertree} shows a schematic representation of a
merger tree illustrating its anatomy. We refer to the halo at the base
of the tree (i.e., the large purple halo at $z=z_0$) as the {\it host
  halo}. For each individual branching point along the tree (one
example is highlighted in Fig.~\ref{fig:mergertree}), the end-product
of the merger event is called the {\it descendant halo}, while the
haloes that merge are called the {\it progenitors}. The {\it main
  progenitor} of a descendant halo is the progenitor that contributes
the most mass. For example, for the branching point highlighted in
Fig.~\ref{fig:mergertree}, the purple halo at $z=z_2$ is the main
progenitor of its descendant at $z=z_1$.  The {\it main branch} of the
merger tree is defined as the branch tracing the main progenitor of
the main progenitor of the main progenitor, etc. (i.e., the branch
connecting the purple haloes). Note, that the main progenitor halo at
redshift $z$ is not necessarily the most massive progenitor at that
redshift. Throughout we shall occasionally refer to the main
progenitor haloes of a given host halo as its zeroth-order
progenitors, while the mass history, $M(z)$, along this branch is
called the {\it mass assembly history} (MAH). Haloes that accrete
directly onto the main branch are called first-order progenitors, or,
after accretion, first-order subhaloes. Similarly, haloes that accrete
directly onto first-order progenitors are called second-order
progenitors, and they end up at $z=z_0$ as second-order subhaloes (or
sub-subhaloes) of the host halo. The same logic is used to define
higher-order progenitors and subhaloes, as illustrated in
Fig.~\ref{fig:mergertree}. Note that with our definition, the mass of
a $n^{\rm th}$-order subhalo {\it includes} the masses of its own
subhaloes (i.e., those of order larger than $n$). Finally, the small shaded
boxes present at each branching reflect the mass accreted by the
descendant halo in the form of smooth accretion (i.e., not part of any
halo) or in the form of progenitor haloes with masses below the mass
resolution of the merger tree. Throughout this paper we shall refer to
this component as {\it smooth accretion}.

\subsection{Merger Trees based on EPS formalism}
\label{sec:EPStrees}

The extended Press-Schechter (EPS) formalism, also known as the
excursion-set formalism, developed by Bond \etal (1991) and Bower
(1991), uses the statistics of Gaussian random fields to compute the
conditional probability $P(M_1,z_1|M_0,z_0) \rmd M_1$, that a
halo of mass $M_0$ at redshift $z_0$ has a progenitor with mass in the
range $[M_1,M_1+\rmd M_1]$ at redshift $z_1 > z_0$.  This conditional
probability function is the basis from which one can construct an
(EPS-based) halo merger tree.

Following Lacey \& Cole (1993), we use the variables $S \equiv
\sigma^2(M)$ and $\omega \equiv \delta_\rmc(z)$ to label mass and
redshift, respectively.  Here $\sigma^2(M)$ is the variance of
the density field, linearly extrapolated to $z=0$ and smoothed with a
sharp $k$-space filter of mass $M$, and $\delta_\rmc(z)$ is the
critical overdensity for collapse at redshift $z$. In the case of
spherical collapse, $\delta_\rmc(z) = 1.686/D(z)$ with $D(z)$ the
linear growth rate normalized to unity at $z=0$.  According to the EPS
formalism, the conditional probability function $P(M_1,z_1|M_0,z_0)$
is given by
\begin{equation}\label{fSC}
P(M_1,z_1|M_0,z_0) = f_{\rm SC}(S_1,\omega_1|S_0,\omega_0) \, \left\vert
{\rmd S_1 \over \rmd M_1} \right\vert\,,
\end{equation}
where $S_i = S(M_i)$, $\omega_i = \omega(z_i)$, and
\begin{equation}\label{multiplicity}
f_{\rm SC}(S_1,\omega_1|S_0,\omega_0) = {1 \over \sqrt{2\pi}} \,
{\Delta\omega \over \Delta S^{3/2}} \, \exp{\left( -{\Delta\omega^2
    \over 2\Delta S} \right)}\,,
\end{equation}
with $\Delta S \equiv S_1 - S_0$ and $\Delta\omega = \omega_1 -
\omega_0$. The progenitor mass function (hereafter PMF) at $z=z_1$
for a host halo of mass $M_0$ at $z_0$ is simply related to the
mass-weighted conditional probability function by
\begin{equation}\label{PMF}
n_{\rm EPS}(M_1,z_1|M_0,z_0) \, \rmd M_1 = {M_0 \over M_1} \, P(M_1,z_1|M_0,z_0) \,
\rmd M_1 \,.
\end{equation}
Note that the PMF $n_{\rm EPS}(M_1,z_1|M_0,z_0)$ is also sometimes
denoted as $\rmd N(M_1,z_1|M_0,z_0)/\rmd M_1$, in that $n_{\rm EPS}(M_1,z_1|M_0,z_0)\rmd M_1$ is the ensemble average number $\rmd N(M_1,z_1|M_0,z_0)$ that a halo of mass $M_0$ at redshift $z_0$ has a progenitor with mass in $[M_1,M_1+\rmd M_1]$ at redshift $z_1 > z_0$. In the case of
ellipsoidal collapse, the same formalism can be used, but with $f_{\rm
  SC}$ replaced by
\begin{eqnarray}\label{fEC}
\lefteqn{
f_{\rm EC}(S_1,\omega_1|S_0,\omega_0) = {A_0 \over \sqrt{2\pi}} \,
{\Delta\omega \over \Delta S^{3/2}} \, 
\exp{\left( -{A_1^2 \over 2}\tilde{S}\right)} \nonumber} \\
& & \times \left\{ \exp\left(-A_3 {\Delta\omega^2 \over 2 \Delta S}\right) 
+ A_2 \tilde{S}^{3/2} \left[1 + 2 A_1 \sqrt{\tilde{S}\over\pi}\right]\right\}\,,
\end{eqnarray}
where $A_0 = 0.8661(1- 0.133\nu_0^{-0.615})$, $A_1 = 0.308
\nu_0^{-0.115}$, $A_2 = 0.0373\nu_0^{-0.115}$, $A_3 = A_0^2 + 2 A_0
A_1 \sqrt{\Delta S \, \tilde{S}}/\Delta\omega$, $\nu_0 =
\omega^2_0/S_0$, and $\tilde{S} = \Delta S/S_0$ (see Zhang \etal
2008a,b for details).

In order to construct an EPS merger tree, one starts from some target
host halo mass, $M_0$, at some redshift $z_0$, and uses the PMF to
draw a set of progenitor masses ${M_1, M_2,...,M_{N_\rmp}}$ at some earlier
time $z_1 = z_0 + \Delta z$, where $\sum_{i=1}^{N_\rmp} M_i = M_0$ in
order to assure mass conservation. The time-step $\Delta z$ used sets
the `temporal resolution' of the merger tree. This procedure is then
repeated for each progenitor with mass $M_i > M_{\rm res}$, thus
advancing `upwards' along the tree. The minimum mass $M_{\rm res}$
sets the `mass resolution' of the merger tree and is typically
expressed as a fraction of the final host mass $M_0$.

There are two problems with this approach. First of all, although EPS
provides the PMF, it does not explicitly specify how to split descendants into progenitors.  
In fact, this can be done using many
different ways, resulting in merger trees with different
statistics. Secondly, the EPS formalism is at best a crude
approximation, and the PMF that it predicts may not be sufficiently
accurate to yield reliable merger trees. We now discuss each of these
two issues in turn.

\subsubsection{The Self-consistency Constraint}
\label{sec:self}

The requirement for mass conservation implies that the probability for
the mass of the $n^{\rm th}$ progenitor of some descendant needs to be
conditional on the masses of the $n-1$ progenitor haloes already drawn.
Unfortunately, these conditional probability functions are not
derivable from the EPS formalism, which results in ambiguity as to how
to partition the descendant mass into progenitor masses. This has
resulted in the construction of a variety of different Monte Carlo
algorithms to construct halo merger trees within the same EPS
framework, i.e., relying on the same $n_{\rm EPS}(M_1,z_1|M_0,z_0)$.

In order to be consistent with EPS, it is crucial that the Monte Carlo
algorithm used to construct the merger trees {\it exactly} reproduces
the EPS conditional mass function $n_{\rm EPS}(M_1,z_1|M_0,z_0)$ for a single
time step $\Delta z = z_1 - z_0$. As shown by Zhang \etal (2008b),
this is a sufficient condition for the algorithm to also reproduce the
$n_{\rm EPS}(M,z|M_0,z_0)$ for {\it any} $z$, regardless of the number, or
width, of intervening time-steps. We shall refer to this as the
self-consistency requirement for the Monte Carlo algorithm.

Several Monte-Carlo merger-tree algorithms rely on the assumption that
in the limit of sufficiently small time-steps, all mergers are binary
in nature (e.g., Cole 1991; Lacey \& Cole 1993; Cole \etal 2000;
Moreno, Giocoli \& Sheth 2008).  Under this assumption it is trivial
to assign the progenitor masses using the constraint of mass
conservation; after drawing the first progenitor mass, $M_1$, from
$n_{\rm EPS}(M_1,z_1|M_0,z_0)$ the mass of the second progenitor is simply $M_2
= M_0 - M_1$. An implicit assumption of this binary method is that the
PMF is symmetric, such that $n_{\rm EPS}(M_1,z_1|M_0,z_0) = n_{\rm
  EPS}(M_0-M_1,z_1|M_0,z_0)$. However, as shown by Sheth \& Pitman
(1997), this is only correct for Poisson initial conditions ($P(k) =
k^n$ with $n=0$). For more relevant cases, such as CDM, $n_{\rm
  EPS}(M_1,z_1|M_0,z_0)$ is (slightly) asymmetric, even in the limit
$\Delta z \rightarrow 0$.\footnote{This also implies that the
  assumption of binary mergers is incorrect, even in the limit $\Delta
  z \rightarrow 0$ (see Neistein \& Dekel 2008b).} Consequently, all
binary methods violate the self-consistency constraint.  Cole \etal
(2000) tried to remedy this by explicitly accounting for accretion of
objects below the mass resolution, $M_{\rm res}$, of the merger
tree. However, as shown in Zhang \etal (2008b) this method still
violates the self-consistency constraint, albeit at a much reduced
level (see \S\ref{sec:PMFs} below).

In order to overcome these problems, several algorithms have been
developed that do not make the implicit assumption of binarity (e.g.,
Kauffmann \& White 1993; Sheth \& Lemson 1999; Somerville \& Kolatt
1999; Zhang \etal 2008b; Neistein \& Dekel 2008b).  Among these, only
the methods of Kauffmann \& White (1993; hereafter KW93), Zhang \etal
(2008b), and Neistein \& Dekel (2008b) fulfill the self-consistency
requirement. The method of Sheth \& Lemson (1999) is only exact for
Poisson initial conditions, while the method of Somerville \& Kolatt
(1999) violates self-consistency, because it discards progenitors
drawn from the conditional mass function that overflow the mass budget
(see \S\ref{sec:SK99} below).

\subsubsection{Beyond Spherical Collapse}
\label{sec:epsplus}

In addition to problems related to the self-consistency constraint,
EPS-based merger trees are also hampered by the fact that EPS is an
approximate theory at best. This implies that the PMF, $n_{\rm
  EPS}(M_1,z_1|M_0,z_0)$, obtained using EPS theory, may not be
sufficiently accurate.  Indeed, comparison with numerical simulations
has shown that the EPS conditional mass function based on the
assumptions of spherical collapse overpredicts (underpredicts) the
number of low-mass (massive) progenitors (Cole \etal 2008). Related to
this is the well-known problem that EPS predicts halo assembly to
occur later than what is found in numerical simulations (e.g., van den
Bosch 2002; Lin, Jing \& Lin 2003; Neistein \etal 2006).

These problems seem to be related to the assumption that halo collapse
is a spherical process. Several studies have shown that assuming
ellipsoidal, rather than spherical, collapse conditions, results in
overall halo mass functions and halo formation times in better
agreement with numerical simulations (e.g., Sheth, Mo \& Tormen 2001;
Hiotelis \& Del Popolo 2006; Giocoli \etal 2007). In the excursion set
approach, the problem of estimating halo abundances reduces to that of
computing the number of time steps a Brownian-motion random walk must
take before it crosses an overdensity barrier. In the spherical collapse (SC)
picture, this barrier has a constant height (i.e., the critical overdensity for collapse is independent of mass scale), allowing one to calculate
the up-crossing statistics for a Gaussian random field
analytically. In the case of ellipsoidal collapse (EC), however, the
constant barrier needs to be replaced with a `moving barrier', i.e., a
barrier that depends on mass scale (see Mo \etal 2010 for
details). Based on the success of EC in predicting the {\it
  unconditional} halo mass function, a number of algorithms have been
developed that use the PMF derived under EC conditions.  Moreno \etal
(2008) approximate the up-crossing barrier as being proportional to
the square root $\sigma(M)$ of the mass variance, in which case the conditional mass
function can be obtained analytically.  Unfortunately, this barrier
shape is different from the one predicted based on EC considerations
(Sheth \& Tormen 2002). In addition, the Moreno \etal merger-tree
algorithm assumes binary mergers, resulting in a violation of the
self-consistency algorithm.  Zhang \etal (2008a) computed the PMF
under ellipsoidal collapse conditions using a more general barrier
shape, and showed that it agrees closely with the exact, but
computationally expensive, method developed by Zhang \& Hui (2006).
Zhang \etal (2008b) then used this EC PMF, given by Eqs.~(\ref{fSC}),
(\ref{PMF}) and (\ref{fEC}), to develop a number of different
merger-tree algorithms, some of which we will discuss and test in this
paper.

Although the EC assumptions generally yields EPS predictions in better
agreement with simulations, significant discrepancies remain (e.g.,
Sheth \& Tormen 2002; Zhang \etal 2008a; Cole \etal 2008).  This has
prompted a number of studies to develop merger tree algorithms that
use progenitor mass functions calibrated to match certain $N$-body
simulations (Neistein \& Dekel 2008a; Cole \etal 2008; Parkinson \etal
2008). These methods basically side-step EPS, but instead inherit all
the problems related to subhalo identification and limitations due to
finite mass resolution discussed above.

\section{Merger Tree Algorithms}
\label{sec:algorithms} 

The main goal of this paper is to assess the performances, compared to
numerical simulations, of a number of merger tree algorithms regarding
a variety of statistics. In this section we briefly describe the
various merger tree algorithms that enter our comparison. These are
the `N-branch method with accretion' method developed by Somerville \& Kolatt
(1999; hereafter SK99), the binary method of Cole et al. (2000;
hereafter C00) and its modification by Parkinson \etal (2008;
hereafter P08), and several of the algorithms suggested by Zhang \etal
(2008b; hereafter Z08).  What follows is a description of how these
different algorithms select progenitor haloes for a single descendant
halo, which constitutes the building block of a merger tree.

\subsection{Somerville \& Kolatt (1999)}
\label{sec:SK99}

Somerville \& Kolatt (1999) developed a merger-tree algorithm that
does not make the assumption of binary mergers. Their `N-branch method
with accretion' allows for an arbitrary number of progenitors per
time-step, and has been widely used, especially in analytical models
for the population of dark matter subhaloes (see Jiang \& van den
Bosch 2014). The algorithm is based on drawing progenitor masses from
the (mass-weighted) conditional mass function.  With each new halo
drawn it is checked whether the sum of the progenitor masses exceeds
the mass of the descendant.  If this is the case the progenitor is
rejected and a new progenitor mass is drawn.  Any progenitor with mass
$M < M_{\rm res}$ is added to the smooth accretion component $M_{\rm
  smooth}$ (i.e., the formation history of these small mass
progenitors is not followed further back in time).  This procedure is
repeated until the total mass left ($M - M_{\rm smooth} - \sum M_i$)
is less than $M_{\rm res}$. This remaining mass is assigned to $M_{\rm
  smooth}$.

\subsection{Cole et al. (2000)}
\label{sec:C00}

The method of Cole \etal (2000; hereafter C00) is an improvement over
the `block model' developed by Cole \etal (1994) and can be described
as a binary method with (fixed) accretion. Similar to SK99, it treats
the mass in progenitors below the mass resolution, $M_{\rm res}$, as
accreted mass. However, unlike SK99, the smooth accretion mass for a
given time step is deterministic, calculated by integrating the
mass-weighted conditional mass function, i.e.,
\begin{equation}\label{MaccC00}
M_{\rm smooth}(z_1 \rightarrow z_0) = 
\int_0^{M_{\rm res}} n_{\rm EPS}(M_1,z_1|M_0,z_0) \, M_1 \, \rmd M_1\,,
\end{equation}
where $M_0$ is the descendant mass. For each branching point, it is
first decided how many progenitors the descendant has by calculating
the mean number of progenitor haloes in the mass range $[M_{\rm
    res},M_0/2]$, given by
\begin{equation}\label{calPC00}
\calP \equiv \int_{M_{\rm res}}^{M_0/2} n_{\rm EPS}(M_1,z_1|M_0,z_0)\, \rmd M_1\,.
\end{equation}
The merger tree time steps, $\Delta\omega$, are chosen such that $\calP
\ll 1$, to ensure that multiple fragmentation is unlikely.  A random
number, $\calR$, generated in the interval $[0,1]$, is used to
determine whether the descendant has one ($\calR > \calP$) or two
progenitors ($\calR \leq \calP$). In the case of a single progenitor,
its mass is $M_1 = M_0 - M_{\rm acc}$. In the case of two progenitors,
one progenitor mass, $M_1$, is drawn from the progenitor mass function
$n_{\rm EPS}(M_1,z_1|M_0,z_0)$ in the range $[M_{\rm res},M_0/2]$, and
the second progenitor is assigned a mass $M_2 = M_0 - M_1 - M_{\rm
  res}$.

\subsection{Parkinson et al. (2008)}
\label{sec:P08}

Parkinson \etal (2008; hereafter P08) modified the C00 algorithm by
using a progenitor mass function tuned to match results from the
Millennium Simulation (Springel \etal 2005), rather than the EPS
progenitor mass function. Specifically, they used the PMF
\begin{equation}
n(M_1,z_1|M_0,z_0) \equiv n_{\rm EPS}(M_1,z_1|M_0,z_0) \, 
G(S_1/S_0,\omega^2_0/S_0)\,,
\end{equation}
where $G(S_1/S_0,\omega^2_0/S_0)$ is a ‘perturbing’ function that is
tuned to match simulation results. P08 adopted the functional form
\begin{equation}
G(S_1/S_0,\omega^2_0/S_0) = G_0 \, \left({S_1 \over S_0}\right)^{\gamma_1}
\, \left({\omega^2_0 \over S_0}\right)^{\gamma_2}\,,
\end{equation}
which has the advantage that the two terms $G_0$ and
$(\omega^2_0/S_0)^{\gamma_2}$ only enter the integral equations for
$M_{\rm smooth}$ and $\calP$ (eqs.~[\ref{MaccC00}]
and~[\ref{calPC00}], respectively) as multiplicative constants.  Using
merger trees constructed from the Millennium Simulation by Cole \etal
(2008), based on Friends-of-Friends (FoF) groups, P08 inferred the
following best-fit values for the free parameters of their perturbing
function: $G_0 = 0.57$, $\gamma_1 = 0.19$ and $\gamma_2 =-0.005$. We
adopt these parameters throughout.

\subsection{Zhang, Fakhouri \& Ma (2008)}
\label{sec:Z08}

Zhang \etal (2008b; hereafter Z08) developed three new algorithms
(called A, B and C) that all, by construction, satisfy the
self-consistency constraint discussed in \S\ref{sec:self}. And each
algorithm can use either the PMF for spherical collapse, based on
Eq.~(\ref{multiplicity}), or that for ellipsoidal collapse based on
Eq.~(\ref{fEC}). In what follows, we only focus on methods A and B,
and refer to them as Z08X[YY], where X is either A or B and YY is
either SC (for spherical collapse) or EC (for ellipsoidal collapse).

Overall, the Z08 algorithms are considerably more involved than any of
the algorithms discussed above. Here we only sketch the rough idea
behind them, and we refer the interested reader to Zhang \etal (2008b)
for details. The Z08 algorithms are similar to that of SK99, in that
they allow for more than two progenitors per time-step. The most
massive progenitor for a branching point is called the primary
progenitor, while all other progenitors are called secondary.  The
main difference between methods A and B is the mass range over which
the primary progenitor is drawn from the PMF: in the case of Method~A,
this mass range is $[M_0/2,M_0]$. In the case of Method~B this mass
range is modified to $[\alpha\,M_0,M_0]$, where $\alpha$ is defined by
\begin{equation}
\int_{\alpha M_0}^{M_0} n_{\rm EPS}(M_1,z_1|M_0,z_0) \, {\rm d}M_1 = 1\,.
\end{equation}

A somewhat unsatisfactory characteristic of the Z08 algorithms is
that, if there are multiple secondary progenitors in a given time
step, they all have identical masses. In addition, in those cases
neither method A nor method B perfectly conserves mass.  However, as
shown in Z08, despite these shortcomings both methods A and B
accurately satisfy the EPS self-consistency constraints, both for the
SC and EC cases.

\section{Putting the Merger Trees to the Test}
\label{sec:comparison}

In our comparison of the various merger tree algorithms described
above (namely, SK99, C00, P08, Z08A[SC], Z08A[EC], Z08B[SC], and Z08B[EC]), we use the following diagnostics: (i) the progenitor mass
function for a tiny time-step, (ii) the mass assembly history of the
main progenitor, (iii) the merger rate per descendant halo, and (iv)
the unevolved subhalo mass function. These diagnostics are chosen
because they have the potential to reveal subtle differences between
the various merger-tree algorithms, and because they have been studied
using high-resolution cosmological $N$-body simulations, which
provides a benchmark for the comparison. In what follows, we discuss
each of these diagnostics in turn.

Throughout what follows we adopt a flat $\Lambda$CDM cosmology with
$\Omega_{\rmm,0} = 0.25$, $\Omega_{\Lambda,0} = 0.75$, $h = H_0/(100
\kmsmpc) = 0.73$ and with initial density fluctuations described by a
Harrison-Zeldovich power spectrum with normalization $\sigma_8 =
0.9$. We use the transfer function of Eisenstein \& Hu (1998) with a
baryonic mass density $\Omega_{\rmb,0} = 0.045$. This is exactly the
same cosmology as that used for the Millennium simulations (Springel
\etal 2005), thus allowing for a direct comparison. We will refer to
this cosmology as the `Millennium cosmology'.

Unless specifically stated otherwise, we always adopt a time step of
$\Delta \omega = 0.002$, independent of which merger tree algorithm we
use. This easily meets all the time-step criteria described in the
original papers, and we have verified that none of our results are
sensitive to this choice for $\Delta \omega$, as long as it doesn't
get significantly larger than $\sim 0.01$. Finally, all merger trees
are constructed using a mass resolution of $M_{\rm res} = 10^{-4}
M_0$, unless specifically stated otherwise.
\begin{figure*}
\centerline{\psfig{figure=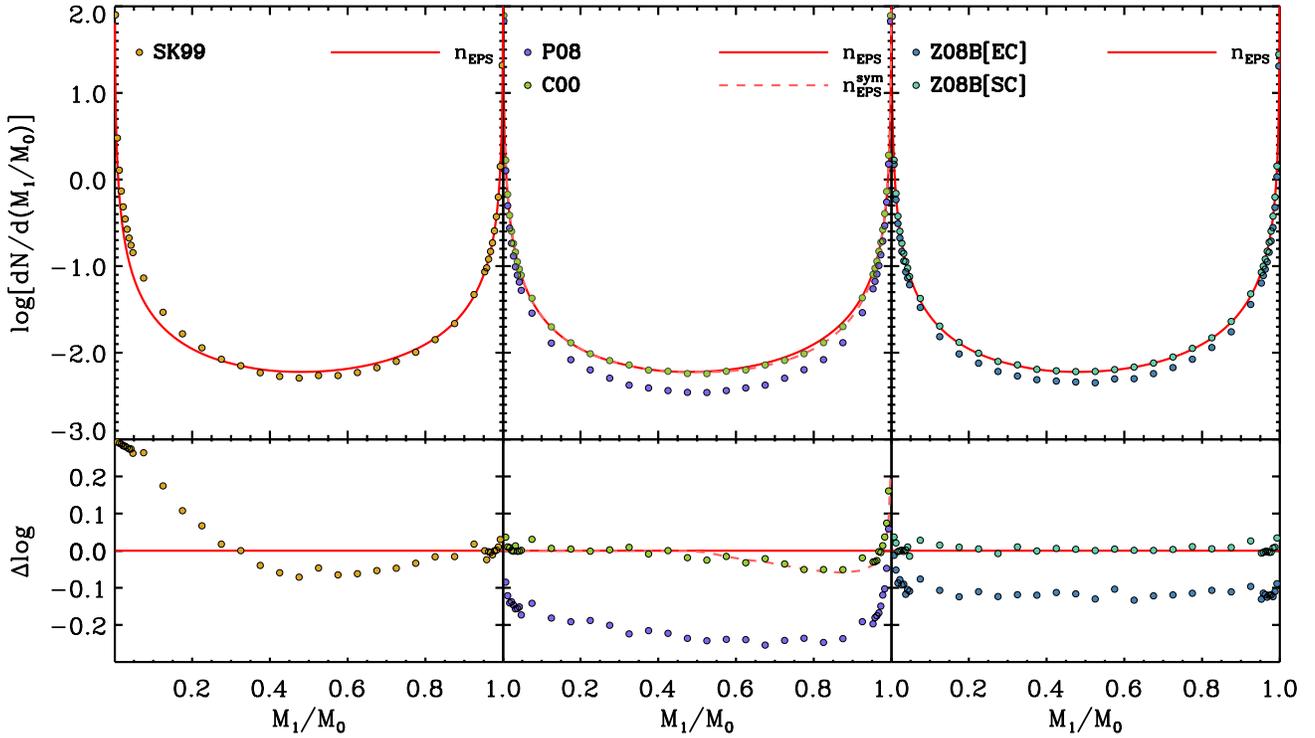,width=\hdsize}}
\caption{Progenitor mass functions obtained from $10^7$ realizations
  of a single time step with $\Delta\omega =0.002$, $M_0 = 10^{13}
  \msunh$ and $z_0=0$ for different merger tree algorithms, as
  indicated (symbols). The solid line indicates the EPS prediction for
  spherical collapse (Eq.~[\ref{multiplicity}] and [\ref{PMF}]), and is shown for
  comparison. The dashed line in the middle panel is the symmetrized
  EPS progenitor mass function of Eq.~(\ref{PMFsymm}), which
  accurately describes the PMF of the C00 algorithm.  Results for the
  Z08A algorithms are not shown, because they are virtually identical
  to those of the Z08B algorithm shown in the right-hand panel.}
\label{Fig:PMFcompare}
\end{figure*}

\subsection{Progenitor Mass Functions}
\label{sec:PMFs}

As a first test of the various merger tree algorithms, we check how
well they perform in terms of the self-consistency test described in
\S\ref{sec:self}.  The symbols in Fig.~\ref{Fig:PMFcompare} indicate
the progenitor mass functions (PMFs) obtained using $10^7$
realizations of a single time step with $\Delta\omega =0.002$, $M_0 =
10^{13} \msunh$ and $z_0=0$ for the various merger tree algorithms
discussed in this paper. The solid line, for comparison, shows the EPS
prediction in the case of spherical collapse (Eq.~[\ref{multiplicity}] and [\ref{PMF}]).
Clearly, SK99 does not meet the self-consistency criterion, in that
the PMF that results from the algorithm doesn't match the EPS
prediction. This is a consequence of the fact that the SK99 algorithm
rejects any progenitor drawn from the EPS PMF that overflows the mass
budget.

The C00 algorithm clearly improves upon this, but it still fails to
meet the self-consistency criterion (at the few percent level) for
$M_\rmp \gta M_0/2$. This is a consequence of the binary assumption
inherent to this algorithm. This is evident from the dashed curve in
the middle panels of Fig.~\ref{Fig:PMFcompare}, which shows the
symmetrized PMF
\begin{eqnarray}\label{PMFsymm}
\lefteqn{
n_{\rm EPS}^{\rm sym}(M_1,z_1|M_0,z_0) \equiv \nonumber} \\
& & \left\{  \begin{array}{ll}
n_{\rm EPS}(M_1,z_1|M_0,z_0)     & \mbox{if $M_1/M_0 \leq 0.5$} \\
n_{\rm EPS}(M_0-M_1,z_1|M_0,z_0) & \mbox{otherwise}
 \end{array} \right. 
\end{eqnarray}
with $n_{\rm EPS}(M_1,z_1|M_0,z_0)$ given by Eq.~(\ref{PMF}).
Clearly, the binary assumption made in the C00 algorithm results in a
PMF that is symmetric with respect to $M_1/M_0 = 0.5$, in disagreement
with EPS (see also Neistein \& Dekel 2008a).

The P08 algorithm results in a PMF that strongly violates the
self-consistency criterion. This is a consequence of the fact that the
P08 algorithm sidesteps EPS by using a `perturbing' function that has
been calibrated such that the resulting PMF is in agreement with that
obtained by Cole \etal (2008) using merger trees extracted from the
Millennium simulation. Hence, the blue dots in the middle panel of
Fig.~\ref{Fig:PMFcompare} are also representative of the PMF in
numerical simulations.

The right-hand panel of Fig.~\ref{Fig:PMFcompare} shows the results
from the Z08B algorithms, both in the case of spherical collapse
(green dots) and ellipsoidal collapse (blue dots). Results for the
Z08A algorithm are not shown, as they are basically indistinguishable
from those of the corresponding Z08B algorithms. Note how the Z08B[SC]
algorithm satisfies the EPS self-consistency criterion to high
accuracy.  Interestingly, the PMF that results from ellipsoidal
collapse conditions falls below that for spherical collapse, very
similar to the PMF of the P08 method. This immediately suggests that
the progenitor mass functions in simulations are more reminiscent of
ellipsoidal collapse conditions than of spherical collapse
conditions. Note, though, that the PMFs of the P08 and Z08B[EC]
methods do differ at the 10 percent level.

\subsection{Mass Assembly Histories}
\label{sec:MAHs}

As discussed in \S\ref{sec:anatomy}, the mass assembly history (MAH)
of a (host) halo is the mass history, $M(z)$, of the halo's {\it main}
progenitor (also called the zeroth-order progenitor). The MAHs of dark
matter haloes have been studied in a large number of papers, using
either the EPS formalism (e.g., Lacey \& Cole 1993; Eisenstein \& Loeb
1996; Nusser \& Sheth 1999; van den Bosch 2002) or $N$-body
simulations (e.g., Wechsler \etal 2002; Zhao \etal 2009; McBride \etal
2009; Fakhouri \etal 2010; Yang \etal 2011; Wu \etal 2013). In this
section we compare the {\it median} MAHs obtained using the different merger tree algorithms to fitting functions obtained from
$N$-body simulations.
\begin{figure*}
\centerline{\psfig{figure=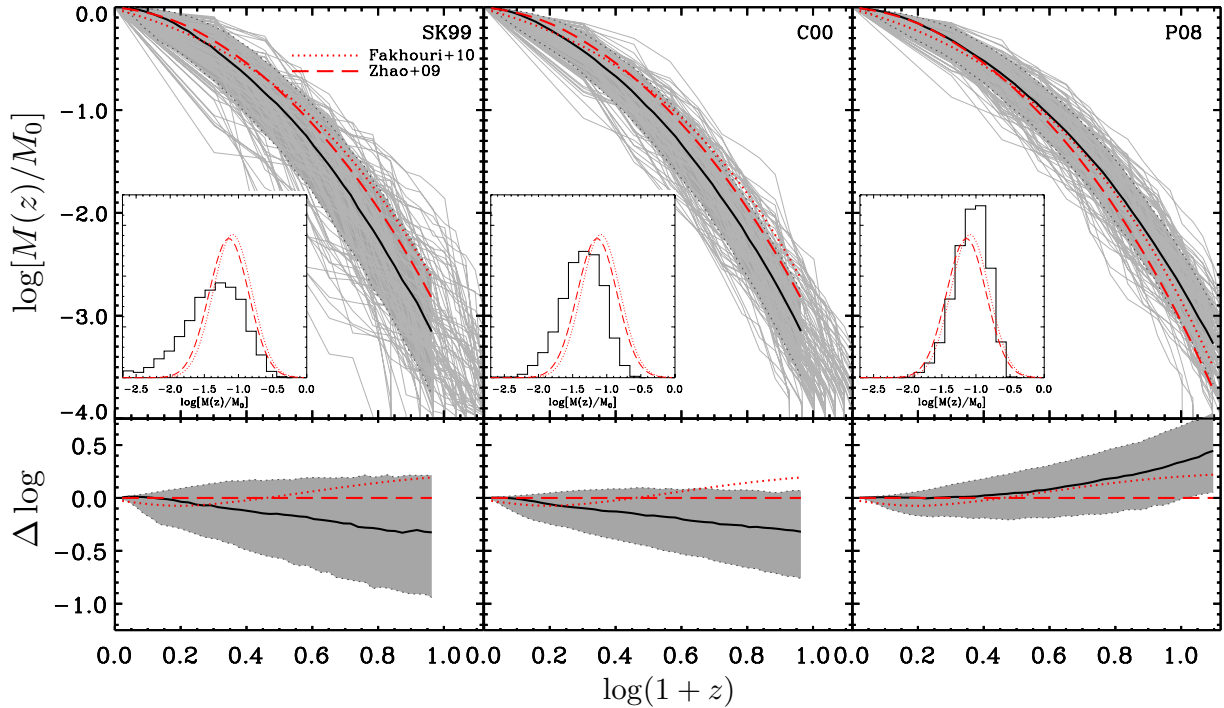,width=0.95\hdsize}}
\caption{Mass assembly histories (MAHs) for present-day host haloes
  with $M_0 = 10^{13} \msunh$ in the Millennium cosmology. The thin,
  gray lines in the upper panels are 100 random realizations obtained
  using the merger tree algorithm indicated in the upper-right corner
  of each panel.  The solid black line indicates the median obtained
  using 2000 MAHs, while the two dotted curves that enclose the shaded
  region indicate the corresponding 68 percentiles of the distribution
  in $\log[M(z)/M_0]$ at fixed redshift.  Note that we only plot the
  median and 68 percentiles up to the redshift where the main
  progenitors of $> 90$\% of all host haloes in consideration can be
  traced (i.e., have masses $M > 10^{-4} M_0$), which can vary
  substantially from one method to the other. For comparison, the red
  curves are the model predictions of the {\it median} from Fakhouri
  \etal (2010, dotted curve) and Zhao \etal (2009, dashed curve), both
  of which are obtained from numerical simulations.  The insets show
  the distributions of $\log[M(z)/M_0]$ at $z=3$ (black histograms),
  as well as two log-normal distributions with the medians taken from
  the Fakhouri \etal and Zhao \etal models, and with the scatter given
  by Eq.~(\ref{scatter}). Finally, the lower panels show the
  differences in $\log[M(z)/M_0]$ with respect to the Zhao \etal
  model.}
\label{fig:MAHsone}
\end{figure*}

Zhao \etal (2009) used a set of cosmological N-body simulations (for
different cosmologies) to study the mass assembly of dark matter
haloes, and generalized from their results a universal model that
predicts the {\it median} MAH for any host halo mass and any cosmology.
Yang \etal (2011) subsequently showed that the scatter in $M(z)/M_0$
at fixed $z$ is well described by a log-normal distribution, with
median given by the Zhao \etal (2009) model and with a dispersion (in
10-based logarithm) given by
\begin{equation}\label{scatter}
\sigma_{\rm MAH} = 0.12 - 0.15 \, \log[\calM(z)/M_0]\,,
\end{equation}
where $\calM(z)$ is the {\it median} main-progenitor mass at $z$.
Because of this log-normal form, it is straightforward to compute the
{\it mean} MAH. After all, for a log-normal distribution the {\it mean}
main progenitor mass, $\langle M \rangle$ is related to the median
according to
\begin{equation}\label{conversion}
\langle M \rangle = \exp\left[ {1 \over 2} \, \left\{ {\rm
    ln}(10)\,\sigma_{\rm MAH}\right\}^2\right] \, \calM \,.
\end{equation}
Thus, assuming a log-normal distribution, we can convert a
{\it median} MAH into a {\it mean} MAH, and vice versa.
\begin{figure*}
\centerline{\psfig{figure=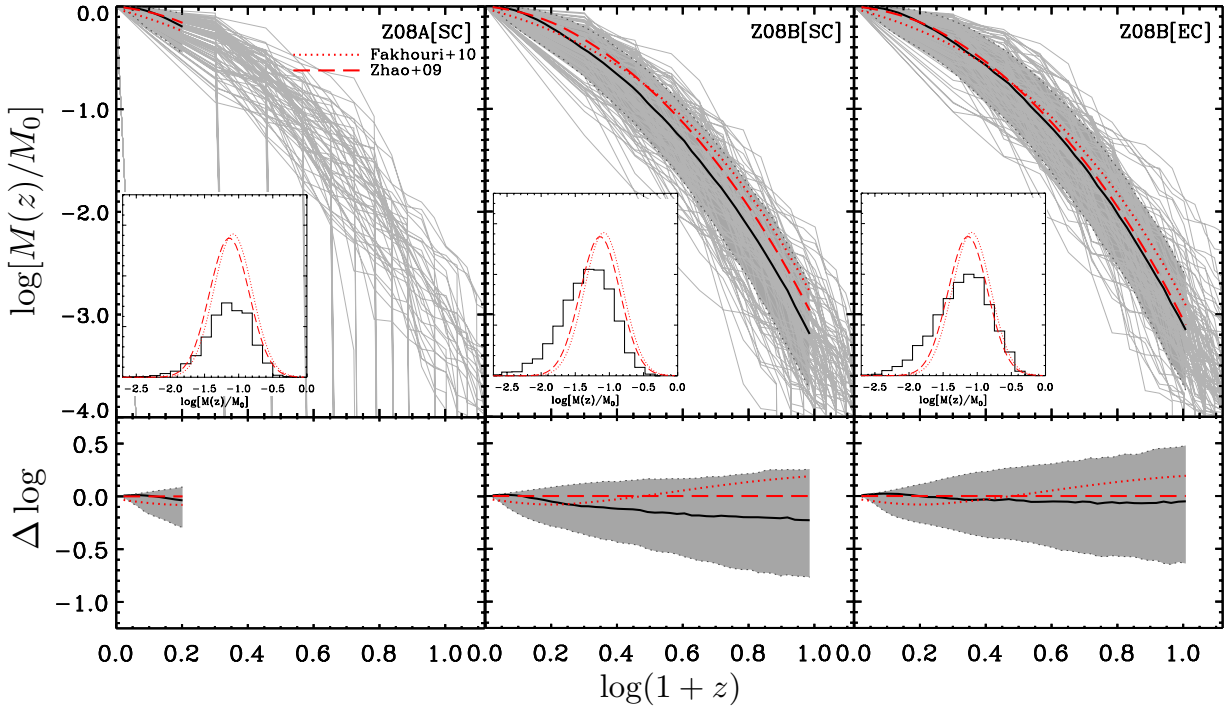,width=0.95\hdsize}}
\caption{Same as Fig.~\ref{fig:MAHsone} but for the merger tree
  algorithms Z08A[SC] (left-hand panels), Z08B[SC] (middle panels) and
  Z08B[EC] (right-hand panels). Note that the histogram in the inset
  for Z08A[SC] has a different normalization than the model curves; this
  reflects the fact that $60\%$ of all MAHs constructed using this
  algorithm have already dropped below the mass resolution of $10^{-4}
  M_0$ by $z=3$.}
\label{fig:MAHstwo}
\end{figure*}

Mcbride \etal (2009) used the Millennium simulation (MS) to study the
mass assembly history and mass growth rate of dark matter
haloes. They provided a fitting formula for the {\it mean} mass growth
rate $\langle \dot{M} \rangle$ as a function of the instantaneous halo
mass $M$ and redshift $z$. In a subsequent paper, Fakhouri \etal
(2010) used a combination of the MS I and II simulations to (slightly)
revise these results, which resulted in a best-fit, mean mass growth
rate
\begin{eqnarray}\label{MassGrowthRate}
\lefteqn{\langle \dot{M} \rangle(M,z) = 46.1\Msun{\rm
  yr}^{-1}\left[\frac{M(z)}{10^{12}\Msun}\right]^{1.1} \, (1 + 1.11\,z) \nonumber} \\ 
& &  \times \sqrt{\Omega_{\rm m,0}(1+z)^3+\Omega_{\Lambda,0}}\,.
\end{eqnarray}
This can be used to model the {\it mean} MAH of a halo by simple
integration. We caution, though, that Eq.~[\ref{MassGrowthRate}]
is only valid for the Millennium cosmology.

Finally, Wu \etal (2013) used the Rhapsody cluster re-simulation
project to study the MAHs of 96 cluster-size haloes of mass $M_0 =
10^{14.8\pm 0.05} h^{-1} \Msun$ at unprecedented resolution (for a
halo sample this size). Their {\it mean} MAH is well fitted by
\begin{equation}
\langle M\rangle(z) = M_0 \, (1+z)^{-1.46} \, {\rm e}^{-0.55 z}\,,
\end{equation}
where $\gamma=0.55$.

In what follows, we compare the simulation results of Zhao \etal
(2009), Fakhouri \etal (2010) and Wu \etal (2013) to the MAHs obtained
using the various merger tree algorithms. Specifically, for a given
cosmology and halo mass, we construct 2000 merger trees from which we
compute the {\it median} MAH. When computing the median, it is
important to take account of the mass resolution of the merger trees
(which we take to be $M_{\rm res} = 10^{-4} M_0$).  Throughout, we
follow Zhao \etal (2009) and only perform our statistical analysis of
the EPS MAHs up to the redshift where the main progenitors of $> 90$\%
of all host haloes in consideration can be traced (i.e., have $M >
10^{-4} M_0$).

Figs.~\ref{fig:MAHsone} and~\ref{fig:MAHstwo} plot the MAHs for haloes
of $M_0 = 10^{13} \msunh$ (at $z_0 = 0$) in the Millennium cosmology.
The thin, gray curves are the realizations for a random subset of 100
MAHs, while the black solid and dotted curves indicate the median and
the 68 percentiles of the distribution of 2000 MAHs. Note that we only
plot this median out to the redshift below which less than 10 percent
of the MAHs have dropped below the mass resolution of the merger tree
($10^{-4} M_0$), which can vary substantially from one method to the
other. For comparison, the red curves are the model predictions, based
on numerical simulations, of Fakhouri \etal (2010) and Zhao \etal
(2009), as indicated. In the case of Fakhouri \etal (2010), we
integrated their model for the mean mass accretion rate
(Eq.~[\ref{MassGrowthRate}]) to obtain the {\it mean} MAH, which we
converted to the {\it median} using Eqs.~(\ref{scatter}) -
(\ref{conversion}). The insets in Figs.~\ref{fig:MAHsone}
and~\ref{fig:MAHstwo} show the distributions of $\log[M(z)/M_0]$ at
$z=3$ (black histograms), as well as two log-normal distributions with
the medians taken from the Fakhouri \etal and Zhao \etal models, and
with the scatter given by Eq.~(\ref{scatter}). Finally, the lower
panels show the differences in $\log[M(z)/M_0]$ with respect to the
Zhao \etal model.
\begin{figure*}
\centerline{\psfig{figure=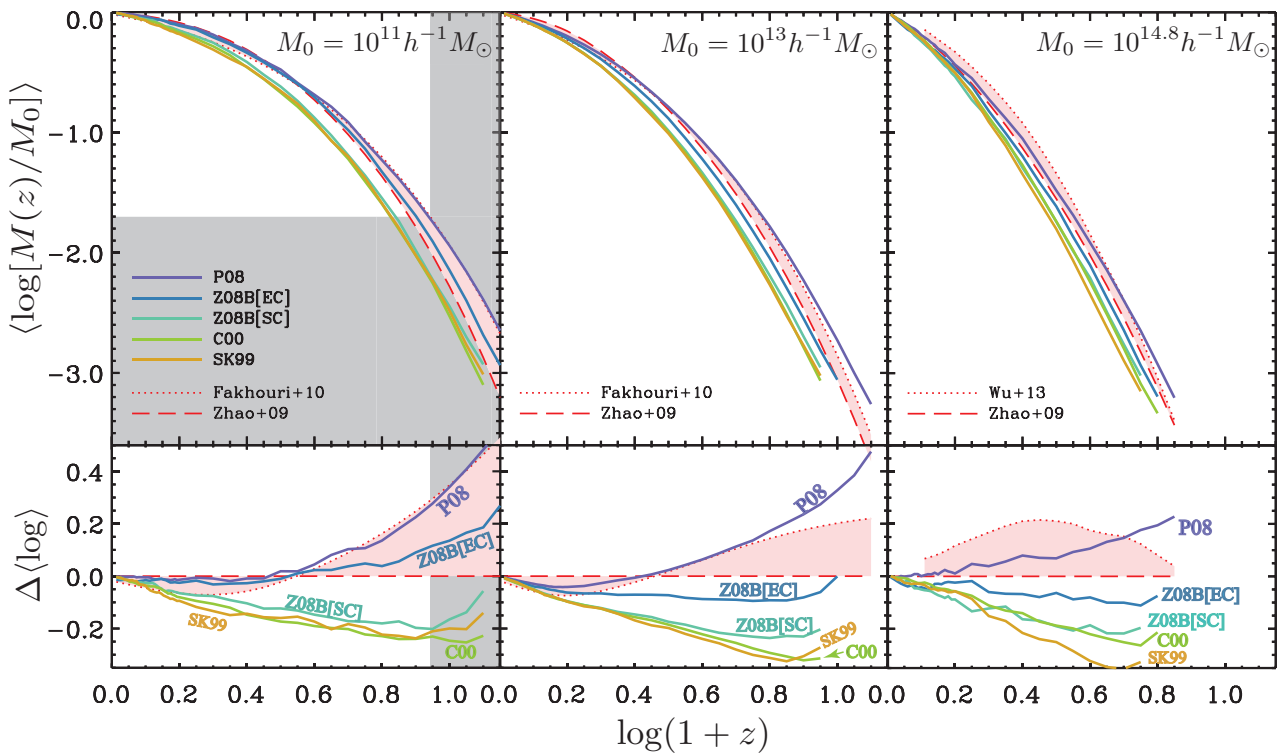,width=\hdsize}}
\caption{Mean $\log[M(z)/M_0]$ as function of redshift for host haloes
  at $z=0$ with masses of $M_0 = 10^{11} \msunh$ (left-hand panel),
  $10^{13} \msunh$ (middle panel), and $10^{14.8} \msunh$ (right-hand
  panel). The solid curves are the results obtained using the SK99,
  C00, P08 and Z08B (both SC and EC) merger tree algorithms, as
  indicated.  In the left and middle panels we have used our fiducial
  `Millennium cosmology', which allows us to compare the MAHs to the
  simulation results of Fakhouri \etal (2010; dotted red curve) and
  Zhao \etal (2009; dashed red curve) model.  The gray-shaded region
  in the left-hand panel roughly marks the mass resolution of the
  simulations used by Fakhouri \etal and Zhao \etal. Hence, the
  simulation results in this region are largely based on
  extrapolation, and have to be considered less reliable. In the
  right-hand panel, in order to facilitate a comparison with the
  results of Wu \etal (2013; dotted red curve), we have adopted the
  `Rhapsody cosmology'. The lower panels show the residuals with
  respect to the Zhao \etal model.}
\label{fig:MAHcompare}
\end{figure*}

The Z08A[SC] method differs from all other methods in that it yields a
large fraction of MAHs that drop below the mass resolution at very low
redshift. In fact, already at $z \simeq 0.4$, more than 10 percent of
the MAHs have dropped below $10^{-4} M_0$. At $z=3$, this fraction has
increased to $40$ percent; for all other methods, zero percent of the
MAHs have dropped out from the sample by $z=3$. Similarly large
`drop-out' fractions are obtained when using method Z08A[EC]. As
discussed in Zhang \etal (2008b), this arises due to a subtlety in how
method A assigns progenitors, and is the main motivation why the
authors considered an alternative; method B. Our results show that
this subtlety yields MAHs that are seriously flawed, and we therefore
no longer consider method Z08A, neither the SC nor the EC version, in
what follows.

Comparing the median MAHs obtained using the various EPS algorithms
with the simulation results, it is clear that the three spherical-collapse-based algorithms, SK99, C00 and Z08B[SC], share a common
feature: they all predict that halo assembly occurs too recent compared
to simulations.  As already discussed in \S\ref{sec:epsplus}, this is
a well-known problem of SC-based EPS. When comparing the scatter in
the MAHs, it is further evident that the SK99 method predicts too much
scatter, while the scatter in the C00 MAHs appears to be in good
agreement with the simulation results. The Z08B[SC] results are
intermediate between those of SK99 and C00.  The ellipsoidal collapse
based method, Z08B[EC], yields a median MAH in excellent agreement
with the simulation results, although the method seems to predict
slightly too much scatter. Finally, the MAHs obtained using the P08
method also are in good agreement with simulations, both in terms of
the median and in terms of the scatter, although there is some
indication that it yields MAHs whose early stages of halo assembly
occur too early.

Fig.~\ref{fig:MAHcompare} plots the mean MAHs obtained using SK99,
C00, P08 and Z08B (both SC and EC) for host haloes at $z=0$ with
masses of $M_0 = 10^{11} \msunh$ (left-hand panel), $10^{13} \msunh$
(middle panel), and $10^{14.8} \msunh$ (right-hand panel).  Note that
contrary to Figs.~\ref{fig:MAHsone} and~\ref{fig:MAHstwo} we now plot
the {\it mean} of $\log[M(z)/M_0]$, which is identical to the {\it
  median} for a log-normal distribution. In the left and middle panels
we have used our fiducial `Millennium cosmology', which allows us to
compare the EPS-based MAHs to the simulation results of Fakhouri \etal
(2010), shown as thick dashed curves. We also show, for the same
cosmology, the predictions based on the Zhao \etal (2009) model.  The
gray-shaded region in the left-hand panel marks the region where the
main progenitor mass $M < 2 \times 10^9 \msunh$. This roughly marks
the mass scale below which the simulations used by Zhao \etal (2009)
and Fakhouri \etal (2010) can no longer reliably resolve the
MAHs. Hence, in the gray region the simulation results are less
reliable and largely based on extrapolation.  Ignoring this region,
the Fakhouri \etal and Zhao \etal models agree roughly at the 0.1-0.2
dex level. In the right-hand panel, in order to facilitate a
comparison with the results of Wu \etal (2013), we adopt the Rhapsody
cosmology (which has lower $\sigma_8$ than the Millennium cosmology,
and a slightly different Hubble parameter). Again, the agreement
between different simulation results, here between Wu \etal and Zhao
et al., is (only) at the level of 0.1-0.2 dex.

A comparison with the MAHs obtained using the various merger tree
algorithms shows once again that the three SC-based algorithms
(SK99, C00 and Z08B[SC]) yield MAHs that systematically fall below
the simulation results. In fact, it is interesting how similar the
average MAHs obtained with these very different methods are.
The P08 and Z08B[EC] algorithms yield MAHs that are in reasonable
agreement with the simulation results; whereas P08 seems to overpredict
$M(z)/M_0$ at early times, Z08B[EC] seems to slightly underpredict
$M(z)/M_0$ for the most massive haloes. 

To summarize, EPS merger trees based on spherical collapse
consistently yields mass assembly histories in which haloes assemble
too late compared to simulations. In terms of halo substructure, this
implies that the SK99, C00 and Z08[SC] algorithms will all
underpredict the accretion redshifts of subhaloes, and are therefore
not well suited to build analytical models for dark matter
substructure or to model satellite galaxies. Both the P08 and Z08B[EC]
algorithms fare much better in that respect. They both yield MAHs in
reasonable agreement with numerical simulations, both in terms of
their median as well as the scatter. The median MAHs predicted by
these two methods are in excellent agreement with each other, and with
the simulation results at low redshifts ($z \lta 2$), but start to
diverge at larger redshifts. At $z = 7$ they typically differ at the
0.3~dex level.  Unfortunately, because of the $\sim 0.2$~dex
discrepancy among the different simulation results, we cannot
significantly prefer one of these two methods over the other.
\begin{figure*}
\centerline{\psfig{figure=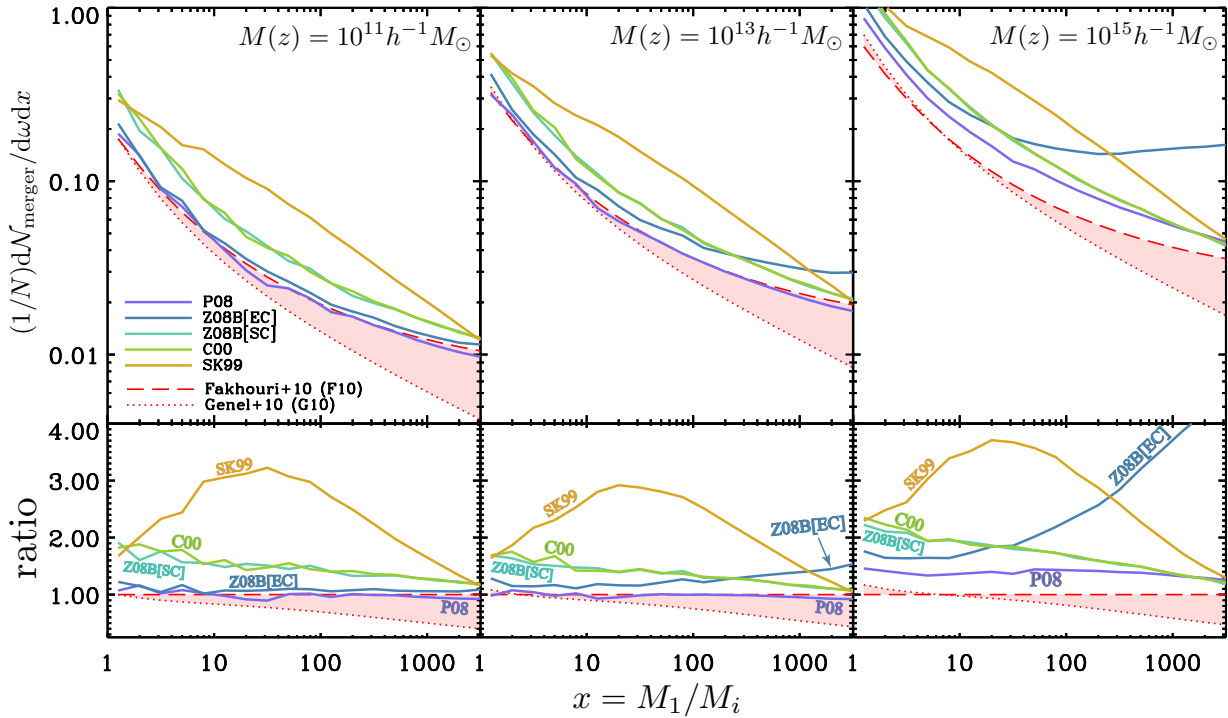,width=0.96\hdsize}}
\caption{Upper panels plot the merger rate per descendant halo, ${\rmd
    \calN_{\rm merger}/{\rmd\omega \,\rmd x}}$, as a function of the
  merger mass ratio $x \equiv M_1/M_i$ ($i=2,3,...$, see text for
  details). Results are shown for descendant haloes at $z=1$ with
  masses of $M(z) = 10^{11} \msunh$ (left-hand panels), $10^{13}
  \msunh$ (middle panels) and $10^{15} \msunh$ (right-hand panels).
  The solid curves are the results obtained using the SK99, C00, P08
  and Z08B (both SC and EC) merger tree algorithms, as indicated,
  using $10,000$ realizations as described in the text.  The dashed
  and dotted red curves are the $N$-body results obtained using the
  Millennium simulation by Fakhouri \etal (2010) and Genel \etal
  (2010), respectively.  The lower panels show the residuals with
  respect to the Fakhouri \etal results.}
\label{fig:MergerRates}
\end{figure*}

\subsection{Merger Rate per Descendent Halo}
\label{sec:MergerRate}

The next diagnostic to consider for our EPS merger tree algorithms is
the merger rate per descendant halo, $(1/N) {\rm d}\calN_{\rm
  merger}/\rmd\omega\,\rmd x$, which characterizes the rate at which
the population of haloes of mass $M = M_1 + M_2$ is created by mergers
between progenitors with a mass ratio $x \equiv M_1/M_2$. Here the
notation is such that $M_i \geq M_{i+1}$, which implies that $x \geq
1$. The quantity $\rmd\calN_{\rm merger}(M,z,x) / \rmd\omega\,\rmd x$
is the number of merger events of mass ratio $x \pm \rmd x/2$ that
result in descendent haloes of mass $M \pm \rmd M/2$ per unit time
interval $\rmd\omega$ at redshift $z$, and $N$ is the number of
descendent haloes of mass $M \pm \rmd M/2$.

Both Fakhouri \etal (2010; hereafter F10) and Genel \etal (2010;
hereafter G10) measured this merger rate per descendant halo from the
Millennium simulations. A problem is that the time steps between
successive outputs of the Millennium simulations are relatively large,
and a descendant halo often has more than two progenitor haloes at the
previous time step. Both F10 and G10 deal with this complication in
the same way: whenever a multiple merger event, consisting of $N_\rmp$
progenitors, occurs they interpret this as a series of $N_\rmp - 1$
binary mergers between $M_1$ and $M_i$ where $i = 2, 3, ...,
N_\rmp$. Although this is not necessarily a proper description of the
true merger history during this time step, this procedure can be
repeated using the EPS formalism, thus allowing for a fair comparison.

Using a combination of the Millennium I and Millennium II simulations,
F10 and G10 found that the merger rate per descendant halo, for the
Millennium cosmology, is well described by;
\begin{equation} \label{MergerRateFitting}
 {1 \over N} {\rmd\calN_{\rm merger} \over \rmd\omega\rmd x}(M,z,x) = 
f(z)\left[ \frac{M(z)}{10^{12}\Msun}\right]^{a_1} x^{a_2} \, e^{a_3 x^{a_4}} ,
\end{equation}
where
\begin{eqnarray}
f(z) = \left\{
  \begin{array}{ll}
0.065                                          & \mbox{(G10)} \\
0.010 \, {\rmd z\over\rmd\omega} \, (1+z)^{a_5} & \mbox{(F10)}
  \end{array} \right. 
\end{eqnarray}
and the best-fit parameters are $(a_1,a_2,a_3,a_4) =
(0.15,-0.3,1.58,-0.5)$ and $(a_1,a_2,a_3,a_4,a_5) =
(0.133,-0.005,3.38,-0.263,0.0993)$ for the G10 and F10 results,
respectively.  Note that the fitting equation of G10 is only valid
over the range $0.5 \lta z \lta 5$. The dashed and dotted curves in
Fig.~\ref{fig:MergerRates} show the merger rates per descendant halo
according to these F10 and G10 fitting formula, as indicated. They are
in reasonable, mutual agreement for $x \lta 10$, but diverge quite
strongly for larger values of the merger mass ratio. In particular,
F10 predicts roughly two times as many minor mergers with $x = 1000$
as G10. This discrepancy mainly arises from the subtle differences in
how these authors extract halo merger trees from their simulation
outputs (see F10 and G10 for details).  As with the MAHs, we are
therefore forced to conclude that different authors obtain halo merger
rates that differ substantially, even when they base their results on
the same simulation.  In what follows, we will simply treat this
discrepancy between the F10 and G10 results as a rough indicator of
the uncertainty on $(1/N) {\rm d}\calN_{\rm merger}/{\rm
  d}\omega\,{\rm d}x$ in simulations.

Using EPS merger trees, it is straightforward to compute the merger
rate per descendent halo.  For a given host halo mass, $M_0$, at a
given redshift, $z_0$, we construct $N = 10,000$ realizations of the
population of progenitor haloes a time $\Delta\omega = 0.002$
earlier. These are used to compute the merger rate per descendant
halo, strictly following the procedure used by F10 and G10 to treat
multiple mergers.  Fig.~\ref{fig:MergerRates} compares the results
obtained using our five remaining merger tree algorithms to the
fitting functions of F10 and G10. All merger rates in this figure are
for a redshift $z_0 = 1$, while different panels correspond to
different host halo masses, as indicated. We have verified that the
results look almost indistinguishable at any other redshift in the
range $0.5 \leq z_0 \leq 5$. We have also verified that the results are not sensitive to the temporal resolution, as we vary $\Delta\omega$ between 0.001 and 0.1, the merger rates at $z_0=1$ change by no more than 10\%.

Upon inspection a number of trends are apparent.  First of all, the
merger rates obtained with the C00 and Z08B[SC] algorithms are
virtually indistinguishable. Both over-predict the rate of major
mergers (mergers with $x \lta 3$) by a factor of about two with
respect to the numerical simulation results of F10 and G10. This
discrepancy becomes smaller for larger $x$, at least when compared to
the F10 fitting function. Interestingly, compared to the G10 fitting
function the C00 and Z08B[SC] merger rates follow almost exactly the
same $x$-dependence, but with a normalization that is a factor $\sim 2$
too high.  A similar trend was noticed by G10. The third spherical
collapse algorithm, SK99, dramatically overpredicts the merger rates
of dark matter haloes compared to both G10 and F10, for all host halo
masses (and at all redshifts). The discrepancy is most pronounced for
mergers with a mass ratio $x \sim 20$, for which SK99 predicts a rate
that is a factor three to four too high. This failure of the SK99
algorithm is a direct manifestation of its failure to satisfy the EPS
self-consistency constraint.

The ellipsoidal-collapse algorithm Z08B[EC] yields an excellent match
to the F10 merger rates for haloes with $M_0 = 10^{11}
\msunh$. However, there is a clear trend that the Z08B[EC] algorithm
starts to overpredicts the merger rates (compared to F10 and G10) for
more massive haloes. This problem is more pronounced for mergers with
a larger mass ratio. For cluster-size host haloes with $M_0 = 10^{15}
\msunh$, and compared to F10, the Z08B[EC] algorithm overpredicts the
major merger rate by a factor 1.7 and that of minor mergers with mass
ratio $x = 1000$ by a factor 3.7. We believe that this problem arises
from the method used to assign halo masses to the secondary
progenitors, which are all assigned the same mass (see \S\ref{sec:Z08}
and Zhang \etal 2008 for details). Finally, the P08 algorithm yields
merger rates that are in excellent agreement with the F10 results. The
only exception seems to be for cluster-size haloes with $M_0 = 10^{15}
\msunh$, where the P08 merger rates are a factor $\sim 1.3$ too high
compared to the F10 fitting function.

To summarize, EPS merger tree algorithms that are based on spherical
collapse overpredict the rate of major mergers by about a factor of
two.  Ellipsoidal collapse seems able to alleviate this
tension. However, the Z08B[EC] implementation of ellipsoidal collapse
has a problem in that it vastly overpredicts the number of minor
mergers for massive host haloes.  Overall, the P08 algorithm yields
merger rates in significantly better agreement with the simulation
results than any of the other merger tree algorithms considered
here. It still overpredicts the merger rates for cluster size host
haloes by about 30 percent, but we emphasize that the disparity in
merger rates obtained by different authors from the same simulation
are of a similar magnitude.
\begin{figure}
\centerline{\psfig{figure=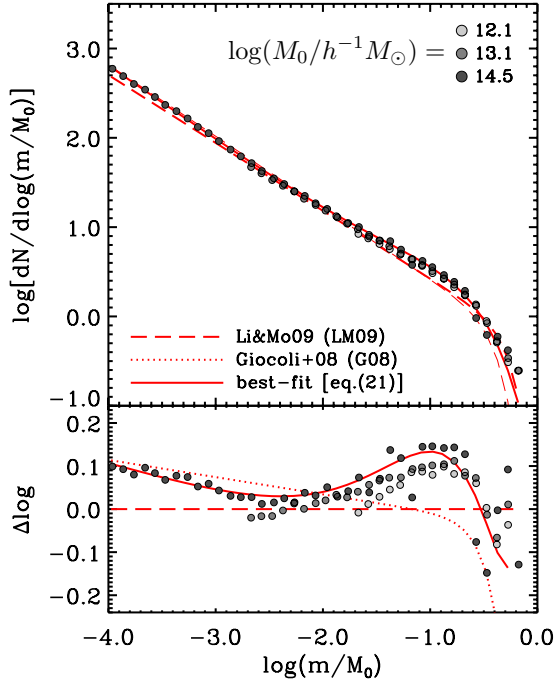,width=0.97\hssize}}
\caption{The unevolved subhalo mass function of first order subhaloes
  (USMF[1]). Symbols are the results obtained by Li \& Mo (2009; LM09)
  using the Millennium simulation, for three different bins in host
  halo mass, as indicated. The dotted and dashed curves are the
  fitting functions obtained by Giocoli \etal (2008; G08) and LM09,
  respectively, while the solid line is the best-fit fitting function
  of the form given by Eq.~(\ref{USMF2}). The lower panel shows the
  residuals with respect to the LM09 fitting function. Note that our
  new fitting function, which has more freedom, better describes the
  `shoulder' around $\log(m/M_0) \simeq -1$.}
\label{fig:USMFsim}
\end{figure}

\subsection{The Unevolved Subhalo Mass Function}
\label{sec:USMF}

The final diagnostic that we consider for testing the various EPS
merger tree algorithms is the unevolved subhalo mass function
(hereafter USMF), ${\rmd N}/\rmd\ln(m/M_0)$, where $m$ is the mass of
the subhalo {\it at accretion}, and $M_0$ is the present-day host halo
mass. 

Using EPS merger trees, van den Bosch, Tormen \& Giocoli (2005)
noticed that the USMF of first-order subhaloes (i.e., only counting
those subhaloes that accrete directly onto the main progenitor) is
universal, in that it doesn't reveal any significant dependence on
either host mass, redshift, or cosmology. This was later confirmed by
Giocoli \etal (2008; hereafter G08) and Li \& Mo (2009; hereafter
LM09) using numerical simulations.  Note, though, that this
universality is only approximate. It is adequate for host haloes with
masses in the range $10^{10} \msunh \lta M_0 \lta 10^{15} \msunh$ in a
$\Lambda$CDM cosmology, but does not necessarily hold for more extreme
halo masses and/or cosmologies.  Indeed, using simulations for
cosmologies with scale-free power spectra, $P(k) \propto k^n$, Yang
\etal (2011) has shown that the USMF depends significantly on the
value of the spectral index $n$. The apparent universality noticed by
van den Bosch \etal (2005), G08 and LM09 arises because the effective
spectral index of the $\Lambda$CDM power spectrum only varies slightly
over the mass range $10^{10} \msunh \leq M_0 \leq 10^{15} \msunh$.
\begin{figure}
\centerline{\psfig{figure=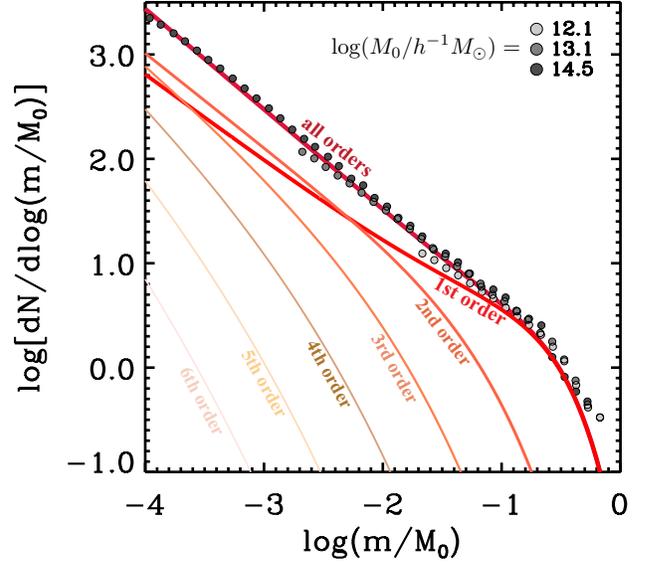,width=0.97\hssize}}
\caption{The unevolved subhalo mass functions (USMF) for different
  order subhaloes, as indicated. The first-order USMF is characterized
  by the fitting function of Eq.~(\ref{USMF2}) with the best-fit
  parameters given in the text, while the higher-order USMFs have been
  computed from this first-order USMF using Eq.~(\ref{recurrence}).
  The data points are the simulation results for {\it all} order
  subhaloes obtained by LM09 using the Millennium simulation, for three
  different bins in host halo mass, as indicated. The solid line
  labeled `all orders' is the corresponding analytical prediction,
  which has been computed by summing the USMFs for order one to four
  (the contribution of higher order USMFs is negligible). The fact
  that this prediction is an excellent match to the data supports the
  notion that the first-order USMF is universal.}
\label{fig:USMFall}
\end{figure}

Using the universality of the USMF of first-order subhaloes, it is
straightforward to compute the USMF of $n^{\rm th}$-order subhaloes,
which is defined as the mass function of $n^{\rm th}$ order subhaloes
at their moment of accretion (i.e., when they transit from being host
haloes to being subhaloes). After all, since subhaloes can themselves
be considered as host haloes at the time of accretion, their
sub-haloes, which are of second-order, are also expected to obey the
universal USMF.  As emphasized in LM09, this implies that
\begin{equation}\label{recurrence}
n_{{\rm un},i}(m|M_0) = 
\int_0^{M_0} n_{{\rm un},1}(m|m_\rma) \, n_{{\rm un},i-1}(m_\rma|M_0) \, \rmd m_\rma\,,
\end{equation}
(for $i=2,3,...$). Here 
\begin{equation}
n_{{\rm un},i}(m|M_0) \equiv {\rmd N \over \rmd m} = {1 \over m} \, 
{\rmd N \over \rmd\ln(m/M_0)}\,,
\end{equation}
is the $i^{\rm th}$-order unevolved subhalo mass function, for which
we will use the shorthand USMF[$i$] in what follows. 
\begin{figure*}
\centerline{\psfig{figure=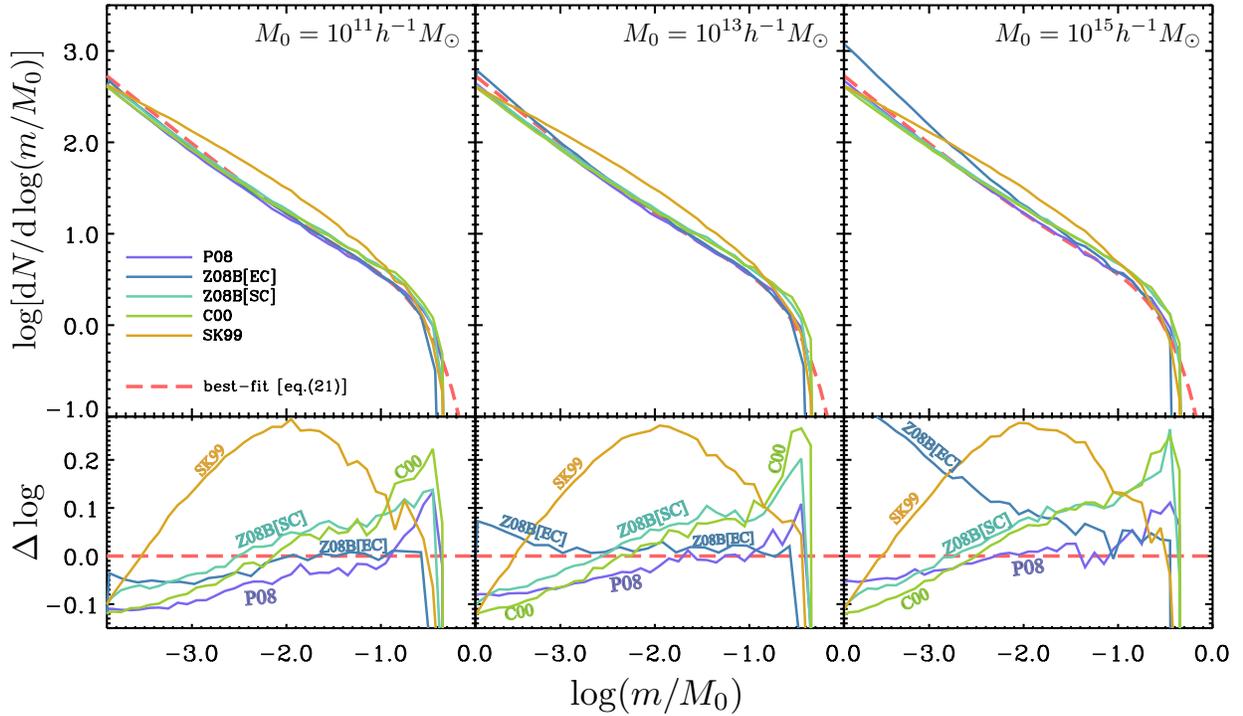,width=0.95\hdsize}}
\caption{The unevolved subhalo mass function of first order subhaloes
  (USMF[1]) for host haloes (at $z=0$) with mass $M_0 = 10^{11}
  \msunh$ (left-hand panels), $10^{13} \msunh$ (middle panels) and
  $10^{15} \msunh$ (right-hand panels).  The solid curves are the
  results obtained using the SK99, C00, P08 and Z08B (both SC and EC)
  merger tree algorithms, as indicated. For comparison, the red,
  dashed line indicates the best-fit representation of the simulation
  results given by Eq.~(\ref{USMF2}) with the best-fit parameters
  given in the text (cf. red, solid curve in Fig.~\ref{fig:USMFsim}).
  The lower panels show the residuals with respect to these simulation
  results.}
\label{fig:USMFcomp}
\end{figure*}

Using the high resolution GIF simulations, G08 found that the USMF[1]
is well fit by
\begin{equation} \label{USMF1}
{\rmd N \over \rmd\ln(m/M_0)} = \gamma \left({m \over M_0}\right)^{\alpha} \,
\exp\left[-\beta\left({m \over M_0}\right)^\zeta\right],
\end{equation}
with best-fit parameters $(\gamma, \alpha, \beta, \zeta) = (0.18,
-0.80, 12.27, 3.00)$.  Note that this implies a total normalization
\begin{eqnarray} \label{USMF1norm}
F_{\rm norm} & \equiv & {1 \over M_0} \int_0^{M_0} m \, {\rmd N \over \rmd m} 
\, \rmd m \nonumber \\
& = & \int_0^1 {\rmd N \over \rmd\ln(m/M_0)} \, \rmd(m/M_0) \simeq 0.735\,.
\end{eqnarray}
The fact that $F_{\rm norm}$ is substantially smaller than unity
implies that dark matter haloes accrete a significant fraction of
their mass `smoothly', either in the form of matter not locked up in
any halo or in the form of haloes with masses below the resolution
limit of the simulation\footnote{Note, though, that the values for
  $F_{\rm norm}$ quoted here are based on extrapolating fitting
  functions for the USMF[1] to subhalo masses well below the mass
  resolution of the simulations.}.

LM09 used the Millennium simulations and found slightly different
best-fit parameters, given by $(\gamma, \alpha, \beta, \zeta) = (0.2,
-0.76, 6.00, 3.20)$, for which $F_{\rm norm} \simeq 0.701$.  The open
circles in Fig.~\ref{fig:USMFsim} are the actual data used by LM09 in
their fitting procedure, for three different bins in host halo mass,
as indicated. The dotted and dashed curves are the best-fit
functions~(\ref{USMF1}) obtained by G08 and LM09, respectively. As is
apparent from the lower panel, showing the residuals, neither is a
good fit to the actual data. In particular, the LM09 data reveals a
clear `shoulder' around the mass scale where the exponential cut-off
kicks in. Since this feature is not captured by the fitting function
of the form~(\ref{USMF1}), we adopt an alternative fitting function
for the USMF, which simply is a linear combination of two components
of the form~(\ref{USMF1}), but with a common exponential-decay part:
\begin{eqnarray} \label{USMF2}
\lefteqn{
{\rmd N \over \rmd\ln(m/M_0)} = \left[\gamma_1
  \left(\frac{m}{M_0}\right)^{\alpha_1} + \gamma_2
  \left(\frac{m}{M_0}\right)^{\alpha_2}\right] \nonumber} \\
& & \times \exp\left[-\beta\left(\frac{m}{M_0}\right)^\zeta\right]\,.
\end{eqnarray}
Using the simulation results of LM09, we obtain the following best-fit
parameters: $(\gamma_1,\alpha_1,\gamma_2, \alpha_2, \beta,
\zeta)=(0.13, -0.83, 1.33,-0.02, 5.67, 1.19)$, which is shown as the
solid line in Fig.~\ref{fig:USMFsim}. Note that this USMF[1] has a
normalization $F_{\rm norm} = 0.8625$, significantly larger than for
the G08 and LM09 fitting functions, which reflects the `extra' mass
under the shoulder. In what follows we will use this new fitting
function for the USMF[1] as the benchmark for our various EPS merger
tree algorithms. As a cautionary remark, we point out that the
simulations used by both G08 and LM09 have a mass resolution of $\sim
2 \times 10^{10} \msunh$, and our fitting function therefore has to be
considered an extrapolation for any $m \lta 2 \times 10^{10} \msunh$.

We have used our best-fit fitting function for the USMF[1] to compute
the USMF for subhaloes of higher-order. The results are shown in
Fig.~\ref{fig:USMFall}, where different curves correspond to USMFs for
different order, as indicated. Note how USMF[2] is higher than USMF[1]
for $m/M_0 \lta 3\times 10^{-2}$, and dominates the total USMF,
defined as the sum of USMF[$i$] for all $i$, over the entire range $-4
\leq m/M_0 \lta -2$.  The contribution to the total USMF due to
subhaloes of order 5 or higher is negligible for all $m/M_0 >
10^{-4}$. The data points in Fig.~\ref{fig:USMFall} are the simulation
results for {\it all} order subhaloes obtained by LM09 using the
Millennium simulation. Note that our prediction for this USMF[all],
which we compute by summing USMF[$i$] for $i=1$ to $4$, is in
excellent agreement with these data\footnote{As shown in LM09, this
  USMF[all] is accurately fit by Eq.(\ref{USMF1}) with $(\gamma,
  \alpha, \beta, \zeta) = (0.22, -0.91, 6.00, 3.00)$.}.

\subsubsection{The Unevolved Subhalo Mass Function of First-Order}
\label{sec:USMFone}

Fig.~\ref{fig:USMFcomp} compares the USFM[1] obtained using the
various EPS merger tree algorithms to our best-fit representation of
the simulation results (dashed line). Results are shown for host
haloes at $z=0$ with $M_0 = 10^{11} \msunh$ (left-hand panel),
$10^{13} \msunh$ (middle panel) and $10^{15} \msunh$ (right-hand
panel). Overall the results are very similar to those for the merger
rate per descendant halo (cf. Fig.~\ref{fig:MergerRates}).  Of all the
merger-tree algorithms considered, SK99 clearly yields the most
discrepant results, overpredicting the USMF[1] by almost 0.3 dex for
subhaloes with $m \sim M_0/100$. The C00 and Z08B[SC] algorithms yield
results that are very similar, and in significantly better agreement
with the simulation results. However, they both underpredict the
USMF[1] for low masses (by about 0.1 dex for $\log(m/M_0) = -4$), and
overpredict it for the most massive subhaloes (by about 0.2 dex for
$\log(m/M_0) = -0.5$). This means that using either of these
algorithms will overpredict the abundance of massive subhaloes
(satellite galaxies) by about 50 percent. The Z08B[EC] algorithm
yields an USMF[1] that is in excellent agreement with the simulation
results for a host halo with $M_0 = 10^{11} \msunh$. However, for more
massive host haloes it starts to overpredict the USMF[1] at the low
mass end. This becomes fairly dramatic for $M_0 = 10^{15} \msunh$,
where the discrepancy exceeds 0.2 dex for $m < M_0/1000$. Finally, as
for the merger rates per descendant halo, the overall best results are
clearly obtained with the P08 algorithm, although it still reveals
deviations from our bench-mark curve of $\lta 0.1$ dex. In particular,
it slightly over(under)-predicts the number of massive (low mass)
subhaloes compared to the results obtained from numerical simulations
\begin{figure*}
\centerline{\psfig{figure=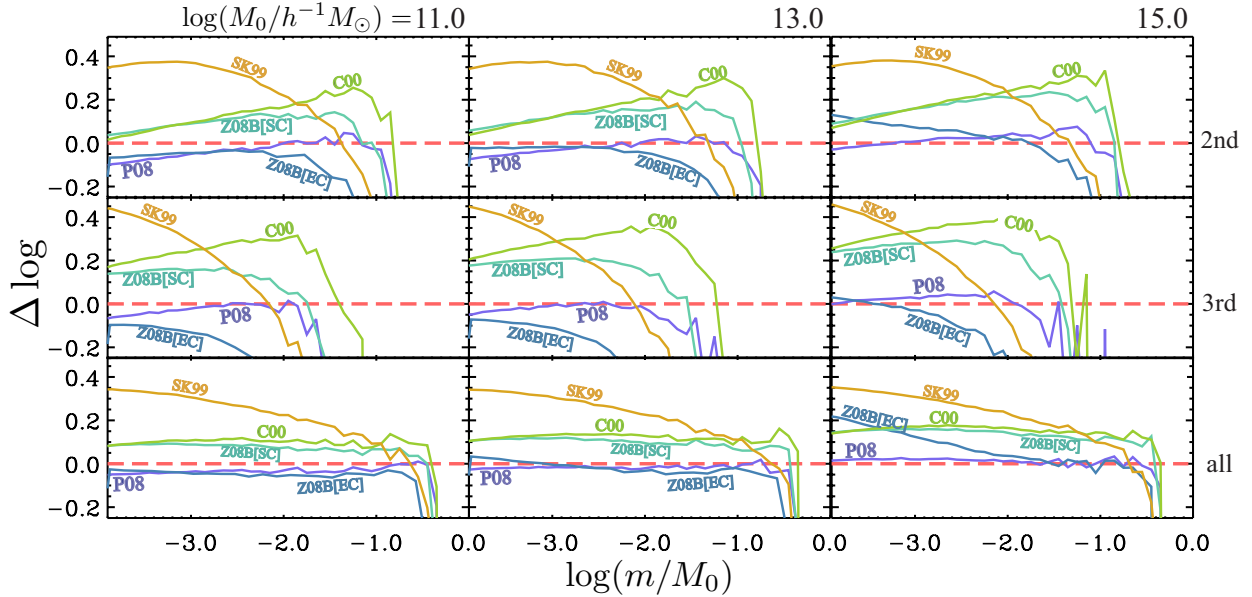,width=0.95\hdsize}}
\caption{Same as the lower panels of Fig.~\ref{fig:USMFcomp} but for
  second-order subhaloes (upper row of panels), third-order subhaloes
  (middle row of panels) and for all orders of subhaloes (lower row of
  panels).  The latter is computed by summing the USMFs of orders one
  to four. See text for a detailed discussion.}
\label{fig:USMFres}
\end{figure*}

\subsubsection{The Unevolved Subhalo Mass Function of Higher Order}
\label{sec:USMFall}

The upper and middle rows of panels in Fig.~\ref{fig:USMFres} show the
residuals of USMF[2] and USMF[3] obtained with the various merger tree
algorithms compared to the analytical predictions based on
Eq.~(\ref{recurrence}). A few trends are apparent. First of all, the
USMFs obtained with SK99, C00 and Z08B[SC] all become progressively
worse for higher order subhaloes, overpredicting the USMF by large
amounts at the low mass end. The USMFs obtained using the Z08B[EC]
method also become progressively worse for higher order, but in the
sense that it starts to underpredict the USMF at the massive end.  The
higher order USMFs obtained using the P08 method, however, become
progressive better with increasing order. The lower panels of
Fig.~\ref{fig:USMFres} show similar residuals but for the USMF of {\it
  all} subhaloes, here defined as the sum of all USMF[$i$] for
$i=1,2,...,4$ (as discussed above, the contribution from higher order
USMFs is negligible for $m/M_0 \geq 10^{-4}$). The P08 predictions are
in excellent agreement with the analytical prediction, which in turn
is in excellent agreement with the simulations results
(cf. Fig.~\ref{fig:USMFall}). The predictions based on the Z08B[EC]
method performs almost equally well for $M_0 < 10^{13} \msunh$, but
over-predicts the abundance of small subhaloes ($m \lta 10^{-2}\,M_0$)
for more massive host haloes. The three SC-based algorithms, SK99, C00
and Z08B[SC] all overpredict the USMF[all] by significant amounts.  In
the case of C00 and Z08B[SC], the offset is roughly independent of
subhalo mass, such that they at least predict the correct power-law
slope for the USMF[all]. The SK99 algorithm, on the other hand,
predicts a power-law slope that is clearly too steep.

\section{Conclusion and Discussion}
\label{sec:Discussion} 

In this paper we have compared and tested a number of different
algorithms for constructing halo merger trees. The diagnostics that we
have used are the progenitor mass functions (PMF), the mass assembly
histories (MAHs), the merger rates per descendant halo, and the
unevolved subhalo mass function (USMF). 

Of all the algorithms tested, the one that fares worst is that of
SK99. The main reason is the strong violation of the self-consistency
constraint (i.e., the progenitor masses drawn fail to sample the
actual PMF), which arises from the fact that the SK99 algorithm
discards progenitor masses drawn from the PMF that overflow the mass
budget (see \S\ref{sec:SK99} for details). In general, the SK99
algorithm yields (i) haloes that assemble too late, (ii) too much
scatter in halo MAHs, (iii) merger rates that are too high by factors
of two to three, and (iv) unevolved subhalo mass functions that are
much too high, especially for subhaloes with a mass at accretion $m
\sim M_0/100$. The latter explains why van den Bosch \etal (2005)
inferred an average subhalo mass loss rate that is too high, as
previously pointed out by Giocoli \etal (2008). It also implies that
other models for dark matter substructure that also used the SK99
algorithm (Taylor \& Babul 2004, 2005; Zentner \& Bullock 2003;
Zentner \etal 2005; Purcell \& Zentner 2012), are likely to suffer
from similar systematic errors.

The various methods introduced by Z08 have the advantage that, by
construction, they accurately satisfy the self-consistency constraint.
This is true for both the spherical and ellipsoidal collapse based
methods. We have tested and compared both methods A and B, and both
for SC and EC. Method A, however, is flawed in that it yields MAHs
that are unrealistic (see the left-hand panel of
Fig.~\ref{fig:MAHstwo}), which is why we haven't considered this
method for any of the other diagnostics. Method B, however, fares much
better. The SC implementation yields MAHs that assemble too late
compared to simulations, and overpredicts the major-merger rate by a
factor of two. These are generic problems for SC-based EPS, and are
therefore shared by the SK99 and C00 algorithms. The Z08B[SC]
algorithm also overpredicts the USMF at the massive end, by about 50
percent. This problem is present for each order of the USMF.  At the
low mass end (i.e., for $m/M_0 \sim 10^{-4}$), the Z08B[SC] method
underpredicts the abundance of first-order subhaloes, but overpredicts
that of higher-order subhaloes.

The ellipsoidal collapse implementation of method B, Z08B[EC], solves
most of these problems. In particular, it yields MAHs, merger rates
and USMFs that are in much better agreement with simulation
results. However, the Z08B[EC] algorithm dramatically overpredicts the
minor merger rate for massive (cluster-size) host haloes. This in turn
results in USMFs for such host haloes that are much too high at the
low mass end. We believe that this failure of the Z08B[EC] algorithm
has its origin in the fact that it assigns all progenitors of a
descendant halo, other than the most massive one, the same mass.

The binary merger method developed by C00 yields halo merger trees
that are extremely similar to those constructed with the Z08B[SC]
algorithm. In particular, it yields MAHs that assemble too late,
overpredicts the major-merger rate by a factor of two, and
overpredicts the USMF of first-order subhaloes at the massive end. The
small (largely insignificant) differences with respect to the Z08B[SC]
algorithm mainly come from the fact that the C00 method violates the
self-consistency constraint, but only for progenitors with a mass more
than half the descendant mass, and only by a few percent. 

Of all the algorithms tested, the one that yields results in closest
agreement with the simulations is the P08 algorithm. It slightly
overpredicts the merger rates for massive descendant haloes, by about
20 percent, and underpredicts the USMF for first-order subhaloes at
the low mass end, by about 15 percent, but given the uncertainties in
the actual simulation results, these discrepancies are barely
significant. The fact that the P08 method yields the best results
should not come entirely as a surprise. After all, P08 draws its
progenitor masses from a PMF that has been modified with respect to
the EPS prediction to match the simulation results presented in Cole
\etal (2008).  Hence, one ought to expect that the P08 algorithm
yields results in better overall agreement with simulations. We
emphasize, though, that even if a method is tuned to reproduce the PMF
of simulations, there is no guarantee that is reproduces any of the
other diagnostics. This requires a merger tree algorithm that
successfully partitions the descendant mass over progenitors, which is
a non-trivial task (as discussed in \S\ref{sec:self}). 

An important (but unavoidable) caveat of the work presented in this
paper is that the benchmarks that we have used to test the various
merger tree algorithms are all based on numerical simulations.  As
discussed in \S\ref{sec:mergertrees}, and as highlighted in this
paper, simulation results carry significant uncertainties that mainly
arise from issues related to identifying haloes and tracking them
across different simulation outputs. The most problematic aspect is
how to properly link (sub)haloes between different snapshots in a
manner that properly accounts for the fact that some sub-haloes are on
orbits that take them outside the host halo's virial radius.  As we
have shown, the discrepancies in average halo mass assembly histories
or merger rates per descendant halo obtained from simulations by
different authors can easily exceed 50 percent, even when they are
based on the same simulation. Clearly, this situation has to improve
if the goal is to build (semi)-analytical models that are accurate to
this level or better. At this point in time, though, taking these
uncertainties into account, we conclude that the accuracy of the P08
merger tree algorithm is not significantly worse that that of the
similations themselves.

As a final remark, we point out that there are several other EPS or
EPS-based algorithms that can be used to construct merger trees (e.g.,
Kauffman \& White 1993; Neistein \& Dekel 2008a,b; Moreno \etal 2008;
method C of Zhang \etal 2008).  Our choice not to include those
methods in this study is simply to keep the project manageable and to
prevent the paper from becoming overly dense. However, we believe it
would be useful to perform similar tests for these alternative methods
as well, and we encourage the community to do so.

\section*{Acknowledgments}

We thank the organizers of the program ``First Galaxies and Faint
Dwarfs: Clues to the Small Scale Structure of Cold Dark Matter'' held
at the Kavli Institute for Theoretical Physics (KITP) in Santa Barbara
for creating a stimulating, interactive environment that started the
research presented in this paper, and the KITP staff for all their
help and hospitality. We also wish to thank Andrew Wetzel and Andrew
Zentner for helpful discussions.  This research was supported in part
by the National Science Foundation under Grant No. PHY11-25915.
 

\label{lastpage}

\end{document}